\documentclass[usenatbib]{mn2e}
\usepackage{graphicx}

\author[Ben Burningham et al.]{Ben Burningham$^1$, Tim Naylor$^1$,
  R. D. Jeffries$^2$ and C. R. Devey$^2$ \\
$^1$ School of Physics, University of Exeter, Stocker Road, Exeter EX4 4QL\\
$^2$ Department of Physics, Keele University, Keele, Staffordshire ST5
  5BG}

\title[On the Nature of Cr 121]{On the Nature of Collinder 121:
  Insights from the Low-Mass Pre-Main Sequence.}

\begin{document}
%
%  These Macros are taken from the AAS TeX macro package version 4.0.
%  Include this file in your LaTeX source only if you are not using
%  the AAS TeX macro package and need to resolve the macro definitions
%  in the BibTeX entries returned by the ADS abstract service.
%
%  If you plan not to use this file to resolve the journal macros
%  rather than the whole AAS TeX macro package, you should save the
%  file as ``aas_macros.sty'' and then include it in your paper by
%  using a construct such as:
%	\documentstyle[11pt,aas_macros]{article}
%
%  For more information on the AASTeX macro package, please see the URL
%	http://www.aas.org/publications/aastex.html
%  For more information about ADS abstract server, please see the URL
%	http://adswww.harvard.edu/ads_abstracts.html
%

% Abbreviations for journals.  The object here is to provide authors
% with convenient shorthands for the most "popular" (often-cited)
% journals; the author can use these markup tags without being concerned
% about the exact form of the journal abbreviation, or its formatting.
% It is up to the keeper of the macros to make sure the macros expand
% to the proper text.  If macro package writers agree to all use the
% same TeX command name, authors only have to remember one thing, and
% the style file will take care of editorial preferences.  This also
% applies when a single journal decides to revamp its abbreviating
% scheme, as happened with the ApJ (Abt 1991).

\def\aj{\rm{AJ}}                   % Astronomical Journal
\def\araa{\rm{ARA\&A}}             % Annual Review of Astron and Astrophys
\def\apj{\rm{ApJ}}                 % Astrophysical Journal
\def\apjl{\rm{ApJ}}                % Astrophysical Journal, Letters
\def\apjs{\rm{ApJS}}               % Astrophysical Journal, Supplement
\def\ao{\rm{Appl.~Opt.}}           % Applied Optics
\def\apss{\rm{Ap\&SS}}             % Astrophysics and Space Science
\def\aap{\rm{A\&A}}                % Astronomy and Astrophysics
\def\aapr{\rm{A\&A~Rev.}}          % Astronomy and Astrophysics Reviews
\def\aaps{\rm{A\&AS}}              % Astronomy and Astrophysics, Supplement
\def\azh{\rm{AZh}}                 % Astronomicheskii Zhurnal
\def\baas{\rm{BAAS}}               % Bulletin of the AAS
\def\jrasc{\rm{JRASC}}             % Journal of the RAS of Canada
\def\memras{\rm{MmRAS}}            % Memoirs of the RAS
\def\mnras{\rm{MNRAS}}             % Monthly Notices of the RAS
\def\pra{\rm{Phys.~Rev.~A}}        % Physical Review A: General Physics
\def\prb{\rm{Phys.~Rev.~B}}        % Physical Review B: Solid State
\def\prc{\rm{Phys.~Rev.~C}}        % Physical Review C
\def\prd{\rm{Phys.~Rev.~D}}        % Physical Review D
\def\pre{\rm{Phys.~Rev.~E}}        % Physical Review E
\def\prl{\rm{Phys.~Rev.~Lett.}}    % Physical Review Letters
\def\pasp{\rm{PASP}}               % Publications of the ASP
\def\pasj{\rm{PASJ}}               % Publications of the ASJ
\def\qjras{\rm{QJRAS}}             % Quarterly Journal of the RAS
\def\skytel{\rm{S\&T}}             % Sky and Telescope
\def\solphys{\rm{Sol.~Phys.}}      % Solar Physics
\def\sovast{\rm{Soviet~Ast.}}      % Soviet Astronomy
\def\ssr{\rm{Space~Sci.~Rev.}}     % Space Science Reviews
\def\zap{\rm{ZAp}}                 % Zeitschrift fuer Astrophysik
\def\nat{\rm{Nature}}              % Nature
\def\iaucirc{\rm{IAU~Circ.}}       % IAU Cirulars
\def\aplett{\rm{Astrophys.~Lett.}} % Astrophysics Letters
\def\apspr{\rm{Astrophys.~Space~Phys.~Res.}}
                % Astrophysics Space Physics Research
\def\bain{\rm{Bull.~Astron.~Inst.~Netherlands}} 
                % Bulletin Astronomical Institute of the Netherlands
\def\fcp{\rm{Fund.~Cosmic~Phys.}}  % Fundamental Cosmic Physics
\def\gca{\rm{Geochim.~Cosmochim.~Acta}}   % Geochimica Cosmochimica Acta
\def\grl{\rm{Geophys.~Res.~Lett.}} % Geophysics Research Letters
\def\jcp{\rm{J.~Chem.~Phys.}}      % Journal of Chemical Physics
\def\jgr{\rm{J.~Geophys.~Res.}}    % Journal of Geophysics Research
\def\jqsrt{\rm{J.~Quant.~Spec.~Radiat.~Transf.}}
                % Journal of Quantitiative Spectroscopy and Radiative Transfer
\def\memsai{\rm{Mem.~Soc.~Astron.~Italiana}}
                % Mem. Societa Astronomica Italiana
\def\nphysa{\rm{Nucl.~Phys.~A}}   % Nuclear Physics A
\def\physrep{\rm{Phys.~Rep.}}   % Physics Reports
\def\physscr{\rm{Phys.~Scr}}   % Physica Scripta
\def\planss{\rm{Planet.~Space~Sci.}}   % Planetary Space Science
\def\procspie{\rm{Proc.~SPIE}}   % Proceedings of the SPIE

\let\astap=\aap
\let\apjlett=\apjl
\let\apjsupp=\apjs
\let\applopt=\ao

\linespread{1.3}
\maketitle

\begin{abstract}

We present a $VI$ photometric catalogue towards the open cluster Cr 121.
XMM-Newton and ROSAT data are used to discover a low-mass pre-main
sequence (PMS) along this sight-line. \citet{zhbbb99} identified Cr 121
as a moving group, using HIPPARCOS data, at a distance of 592 pc. We reject the
scenario that these low-mass PMS stars are associated with that association.
By considering the higher mass main sequence stellar
membership of the groups along this sight-line, the density of low-mass
PMS stars and their age spread we argue that the
low-mass PMS stars are associated with a young, compact cluster at a
distance of 1050 pc. 
 This is consistent with the original description of Cr 121 \citep{c31}, and we
argue that this distant compact cluster should retain its original
designation. 
 The moving group detected by \citet{zhbbb99}
resembles a foreground association and we agree with \citet{e81} that
this should be called CMa OB2.

This study demonstrates that although the \citet{zhbbb99} census of OB
associations is an invaluable resource for studying local star
formation, it must be interpreted in the context other data when
considering structure over distances of the same order as the limits
of the Hipparcos parallaxes.

\vspace{1.5cm}

\end{abstract}

\section{Introduction}
Collinder 121 (Cr 121) is listed in the New List of OB Associations
\citep{me95} as an OB association at a distance of 540 pc.  
However, since its first identification as an open cluster of about 1$^{\circ}$
diameter by \citet{c31}, Cr 121 has had its
membership re-assessed a number of times.
In this paper we use the low-mass pre-main sequence (PMS) to  resolve
the long standing controversy over the nature of Cr 121, and
demonstrate the power of the using the low-mass PMS to pick apart
structure in OB associations.  We also illustrate the difficulties
associated with using
proper motion and parallax data of limited range in isolation.
   
The cluster was first discovered as a group of 18-20 main-sequence
B stars found within a 40x60 arcminute box.  Collinder (1931) used the
cluster diameter, along with mean separation of stars within it, to
derive a distance of 1260 pc. 
  \citet{f67} studied the cluster using UBV photometry and extended
  the membership to 40 stars brighter than $V=7$ within a
  10$^{\circ}$x10$^{\circ}$ box.  
This group was characterised as an OB
association and the position of the zero age main sequence (ZAMS)
indicated a distance of 630 pc. Already, two very different structures
had been described by different authors.
 \citet{e81} suggested that
the original compact group and the larger OB association may be
distinct.  Using intermediate band and H$\beta$ photometry
\citet{e81} placed the compact group, which he referred to as Cr 121,
at a distance of 1.17 kpc at an age of 1.5 Myrs.  The more diffuse
group, which Eggen referred to as CMa OB2, was found to be at a
distance of 740 pc.

More recently emphasis has been placed on the use of common motion
criteria for defining cluster and association memberships.
\citet{zhbbb99} used Hipparcos proper motion and parallax data to
select a moving group in the direction of Cr 121.  They selected 103
stars in a region 13$^{\circ}$x16$^{\circ}$ in extent, and find a mean
distance of 592 $\pm$28 pc.  The presence of an O-star, a Wolf-Rayet
(WR) star and early type B stars in the membership list led to an
estimated age of $\sim$5 Myr.  
Since stars were selected from a large region of sky compared to the
original boundaries of the distant interpretation of Cr 121, and the
fact the Hipparcos parallaxes become unreliable at about 1 kpc, this
selection method is biased to detecting a larger, more diffuse
foreground association rather than the distant compact cluster
of \citet{c31}.  If both groups were present, as suggested by
\citet{e81}, only the foreground association would be detected as this would
represent a large number of stars with common motion at a similar
distance, within the reliable domain of Hipparcos parallaxes.
\citet{dla2001} used data from the Tycho2 catalogue to re-determine
the mean proper motion of Cr 121, amongst other clusters.  This study was
restricted to open clusters within 1 kpc, so again was biased to only
detecting the foreground group, and would have been insensitive to any
more distant cluster along the same line of sight.  Apparently working
within the assumption of the \citet{zhbbb99} view of Cr 121 and the
\citet{dla2001} membership list, \citet{daml2002} determine the
distance to Cr 121 as 470 pc with and age of 11 Myr.

\citet{k2000} carried out Str\"omgren and H$\beta$ photometry of bright
B stars within a 5$^{\circ}$ radius of the classical centre ({\it l,b})
= (235.4$^\circ$,--10.4$^\circ$) of Cr 121.  This revealed a group
of stars at 660-730 pc, with characteristics similar to an OB
association, and a more compact group of stars at a distance of 1085
pc, which she calls Cr 121.  This would indicate that the original cluster
Cr 121 is situated at a distance of over 1 kpc, whilst the Cr 121 selected
by \citet{zhbbb99} is a less distant OB association.

Inspection of the ROSAT Position Sensitive Proportional Counter (PSPC)
catalogue revealed a number of X-ray point sources in the vicinity of
WR 6, near the center of the region studied by \citet{k2000} and the
original center of Collinder's cluster.  WR 6 is
listed as a member of Cr 121 by \citet{zhbbb99}, but not by
\citet{k2000}.  The Hipparcos parallax for WR 6 of $\pi$ =
1.74$\pm$0.76 mas, puts it at a distance of 575 pc.

Low-mass PMS stars are
expected to be X-ray bright due to high levels of coronal activity, as such,
X-ray data has been used as a diagnostic tool for identifying low-mass
PMS stars \citep[e.g.][]{nf99,pjnthk2000}. 
 The presence of low-mass PMS stars would
allow us to fit isochrones to the PMS in a colour-magnitude diagram
(CMD), constraining the distance and age of the PMS.
With this in mind we have obtained deep $VI$ imaging
photometry of a region 20' in radius, covering the central region of
the ROSAT PSPC field of view (FoV) for the pointings carried out
towards WR 6.  Since the optical data were obtained, XMM-Newton data
for this region of sky became available, allowing us to perform a more
sensitive X-ray selection of PMS stars.

\begin{figure}
\includegraphics[height=375pt,width=275pt, angle=90]{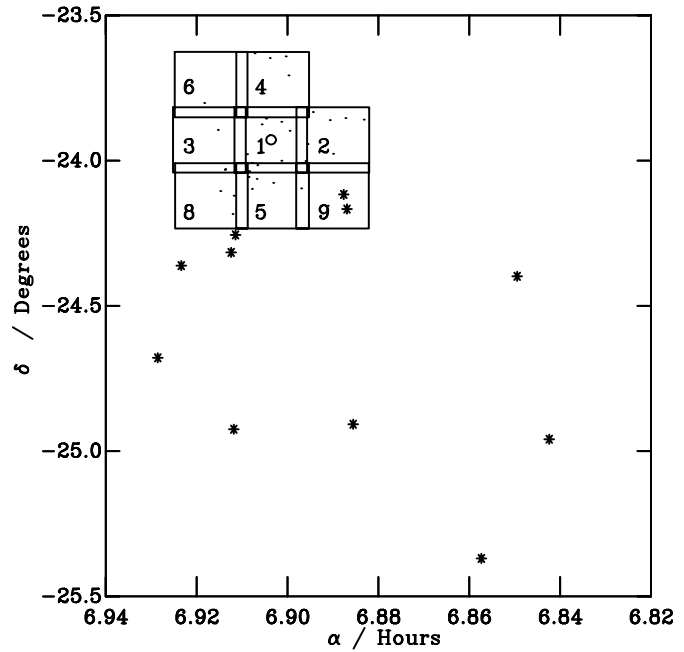}
\caption{The Layout of the fields observed towards WR 6.  The square
  boxes outline the 13.5'x 13.5' CTIO 0.9m fields of view.  The 11 B
  stars listed as members of the more distant association described by
  \citet{k2000} are marked as asterisks.  WR 6 is marked with a
  circle.  The PMS objects discovered here are marked with dots.}
\label{fig:wr6fovs}
\end{figure}

The layout of our optical survey is shown in
figure~\ref{fig:wr6fovs}, along with location of the B stars
identified as being members of the more distant group by and WR 6. 
Although our survey region coincides with only two of the B
stars identified by \citet{k2000} as members of the more distant
cluster, and lies towards the edge of the region enclosed by them, we
would still expect to find a significant number of low-mass PMS stars
from that group within our survey.  Studies of other young clusters
and OB associations have always revealed O and B stars located within a sea
of low-mass PMS stars, not low-mass PMS stars just within a region enclosed
by higher mass stars \citep[e.g.][]{dm2001a,pnjd2003}.

\section{Optical Data}

\subsection{Observations}

The $VI$ CCD imaging of 8 13.5' FoVs centred on WR 6 was
carried out using the 0.9m telescope at the Cerro Tololo Interamerican
Observatory (CTIO) in Chile on the night of 12/13 February 2002.

The FoVs were selected to cover the central region of the ROSAT PSPC
pointings.  Each field was observed with short (10, 6 seconds) and
long exposures (300, 180 seconds) in the $V$ and $I$ bands in
photometric conditions.  The short
exposures were aimed at capturing the brighter stars in each field,
which would be saturated in the long exposures.  To reduce the number
of cosmic ray hits in any one image the long exposures were split into
three parts, of 100s in V and 60s in I.  The fields were overlapped
with each other to provide a check for the internal consistency of the
photometry (see section 2.2).
A number of \citet{l92}
standard star fields, containing standards with -0.6$\leq V-I \leq$5.8,
were also observed at several times during the night.  This ensured
adequate calibration over the colour range of interest.

\subsection{Data Reduction and Optimal Photometry}

The data were reduced, and photometry was performed using the optimal
extraction algorithm originally described by \citet{n98} with its
application for constructing CMDs described in depth by \citet{ntjpdt2002}.
 The standard star fields were reduced using the same algorithm. The
instrumental magnitudes and colours for 116 standard star observations
were compared to those given by \citet{l92} to yield zero-points,
colour terms and extinction coefficients.  We found that an
additional, magnitude independent, uncertainty of 0.02 mag was
required to give a $\chi^2$ of 116 for the $V$ band calibration. No
such adjustment was required for $V-I$.

Each image was bias subtracted and flatfield corrected, using files
constructed from twilight sky flatfield images and bias frames taken
at the beginning of the same evening as the observations.  The
I-band images were combined (after determining their offsets) before
using the deep co-added image for object detection.  Optimal
photometry was then carried out on the list of
stars produced by the detection software to produce an optical
catalogue for each FoV.  Although the I-band images were combined for
object detection, photometry was carried out on the individual
frames.  This ensures that good signal-to-noise in one frame is not
swamped by poor signal-to-noise in another when the images are
combined.  The individual measurements of each star are combined
later, weighting each measurement according to its signal-to-noise.

An astrometric solution was obtained through comparison of the optical
catalogues with SuperCosmos catalogues \citep{supercos1}  of the same
FoVs.  
The RMS of the 6-coefficient fit were approximately 0.2 arcsec.
The resulting catalogues were then combined to produce an optical
catalogue for the entire region covered by our survey, listing
a position, a $V$ magnitude, a $V-I$ colour and a data quality flag
\citep[see][]{ntjpdt2002} for 26104 stars.  
The overlap regions between the fields were used to assess the
internal consistency of the optimal photometry, which was found to be
consistent at the 0.005 mag level.
 The final catalogue given as
table 1, and is available via the CDS online database.  
The reduction carried out
here differs from that described by \citet{ntjpdt2002} in that an
additional data quality flag has been introduced, and the flags are
now indicated by character values rather than integers, since this
gives greater flexibility for the format.  There is also an entirely
new flag, `I', which indicates that the star's photometry is unreliable due to
ill-determined sky.  The threshold for this flag to be invoked is when
the fit to the sky histogram for the star has a reduced $\chi^2 > 3$.  
The resulting CMD is shown below in figure~\ref{fig:cmd}. 
A population of PMS stars is clearly visible on the CMD of the optical
catalogue, lying redward of the background contamination.

\section{X-Ray Data}

\subsection{ROSAT Data}

Nine pointings of the ROSAT PSPC were made towards WR 6 between 1991
and 1992 and total 28 ks.  The data set for each pointing was retrieved
from the public archives via the LEDAS website
(http://ledas-www.star.le.ac.uk/rosat/) hosted by Leicester
University.  The data were reduced using Asterix version 2.3-b1, a
suite of X-ray data reduction programs available from the University
of Birmingham (http://www.sr.bham.ac.uk/asterix).  The data were
sorted into 20 arcminute radius image cylinders (spectral data was
retained at this stage), excluding events recorded during periods of
high background or poor aspect quality and selecting corrected pulse
height (PHA) channels 11 to 201.  The resulting images were then
visually inspected and obvious sources masked in duplicate images, which
were then used to create a background image cylinder for each pointing.
These were then used,
along with other calibration files supplied with the data, to create a
background model for each pointing.  These, and the unmasked image
cylinders were then projected to create 2-d images and background
models, which were used for source detection using Cash-statistic
maximisation in the PSS algorithm \citep{sav2000}.  The resulting
source lists were then used to further mask sources in the background
image cylinders to produce refined background models.  The refined background
models and the images were then co-added.  This resulted in an image
for source detection with an effective exposure time of 25.8 ks.  This
image was then searched for sources using PSS
returning 39 X-ray sources.  The archival release of XMM-Newton
observations of this region of sky means that much of the data in this
source list has been superseded in sensitivity and positional
accuracy.  As a result we retrieved the XMM-Newton data and removed
from the ROSAT PSPC source list those objects that appeared in both
sets of data.

\subsection{XMM-Newton Data}

Data from two XMM-Newton observations using the European Photon
Imaging Camera (EPIC) were used to make the X-ray
source catalogue. One directed at WR 6 with 12864 second total
integration time during revolution 346 \citep{szgs2002},
and another of 52496 seconds directed at the north-west quadrant of
the ring nebula S308 carried out during revolution 343 (unpublished).
\citet{szgs2002} investigated the possible existence of a close
companion to WR 6.  They make note of one other source in their FoV,
57.4'' south of WR 6, which they fail to correlate with a SIMBAD
counterpart.  This source correlates with a star in our photometric
PMS selection (field 1, star 73, see table~\ref{table:xraypms_phot}).  
Other than this they do not discuss the other sources in the FoV.
The final EPIC source lists
were retrieved from the ESA XMM-Newton Science Data Archive
(http://xmm.vilspa.esa.es/) and filtered to remove sources with
detection likelihood values of less than 20.  This was to allow for an
error in the calculation of detection likelihoods by the XMM-SAS
pipeline, in versions 5.4.1 and earlier (see XMM-Newton news \#29).
The two source lists were then cross-correlated with one another to
remove duplicates.
In each case of a duplicate
source the observation with best signal to noise was used.     
The source list was then further restricted to cover just the region
covered by the optical survey.  The resulting source list contained
138 sources, and is given as table 2, which is available via the CDS
online database. Listed for each source are the XMM-Newton source number, right
ascension, declination, error in position, count-rate and error in
count-rate. For those X-ray sources that correlate with an optical
counterpart (see section 4) the field number and star index are also listed.

Of the 39 ROSAT PSPC sources, 13 lay within the optical
survey region and did not appear in the XMM-Newton source list.  We
will refer to this set of sources as the reduced ROSAT PSPC source list.   
This source list is given as table 3, and is available via the CDS
database.  For each source we list and index number, right ascension,
declination and the flux.  Again, for those sources that correlate with
an optical counterpart we also list the field number and star index.

\section{Identification of X-Ray Sources}

\subsection{Cross-correlation of optical and X-ray source ists}
\begin{figure}
\includegraphics[height=375pt,width=275pt, angle=90]{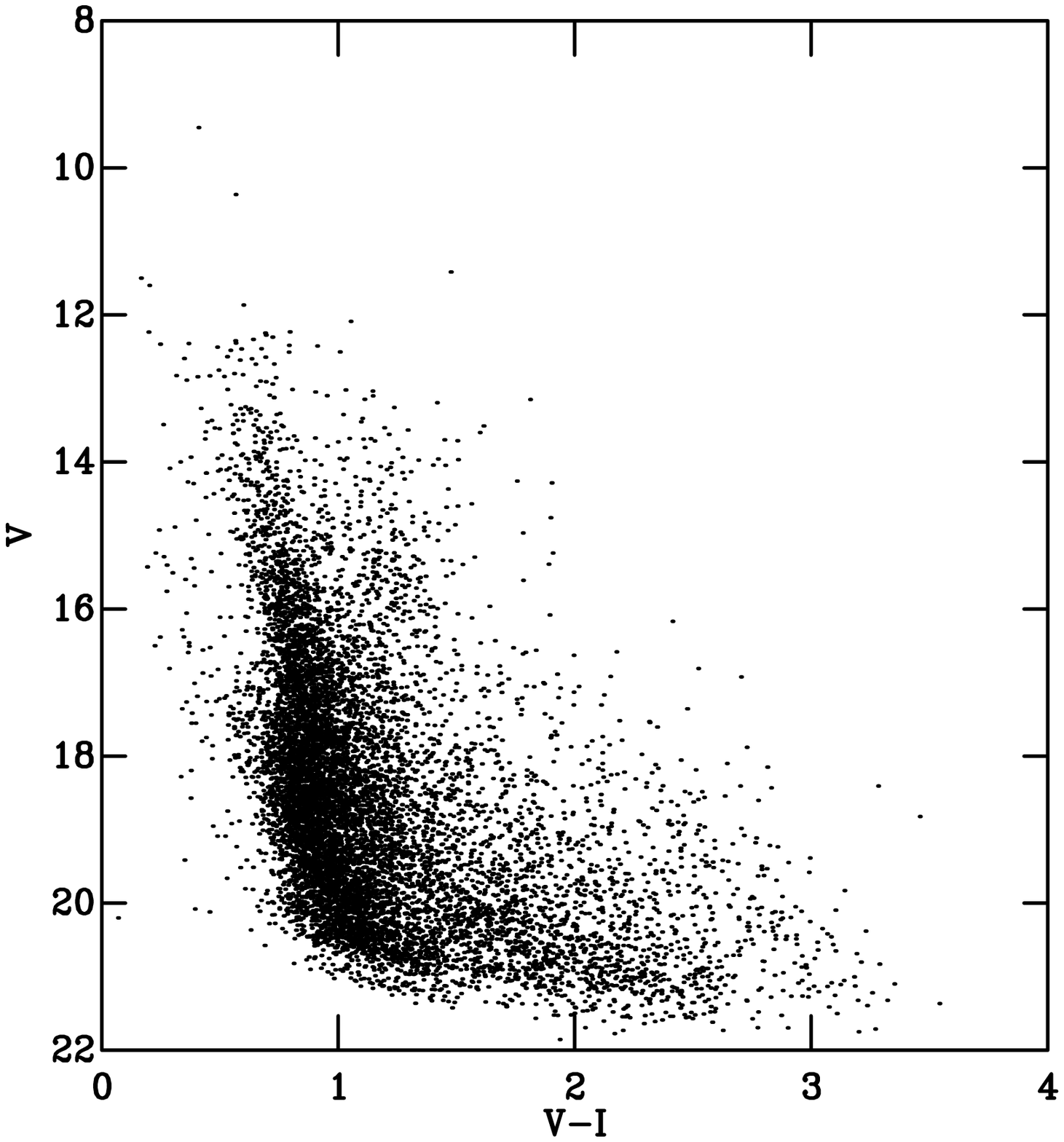}
\caption{The CMD for the optical survey region for all unflagged
  objects with signal-to-noise $>$ 10 in V and V-I.}
\label{fig:cmd}

\includegraphics[height=375pt,width=275pt, angle=90]{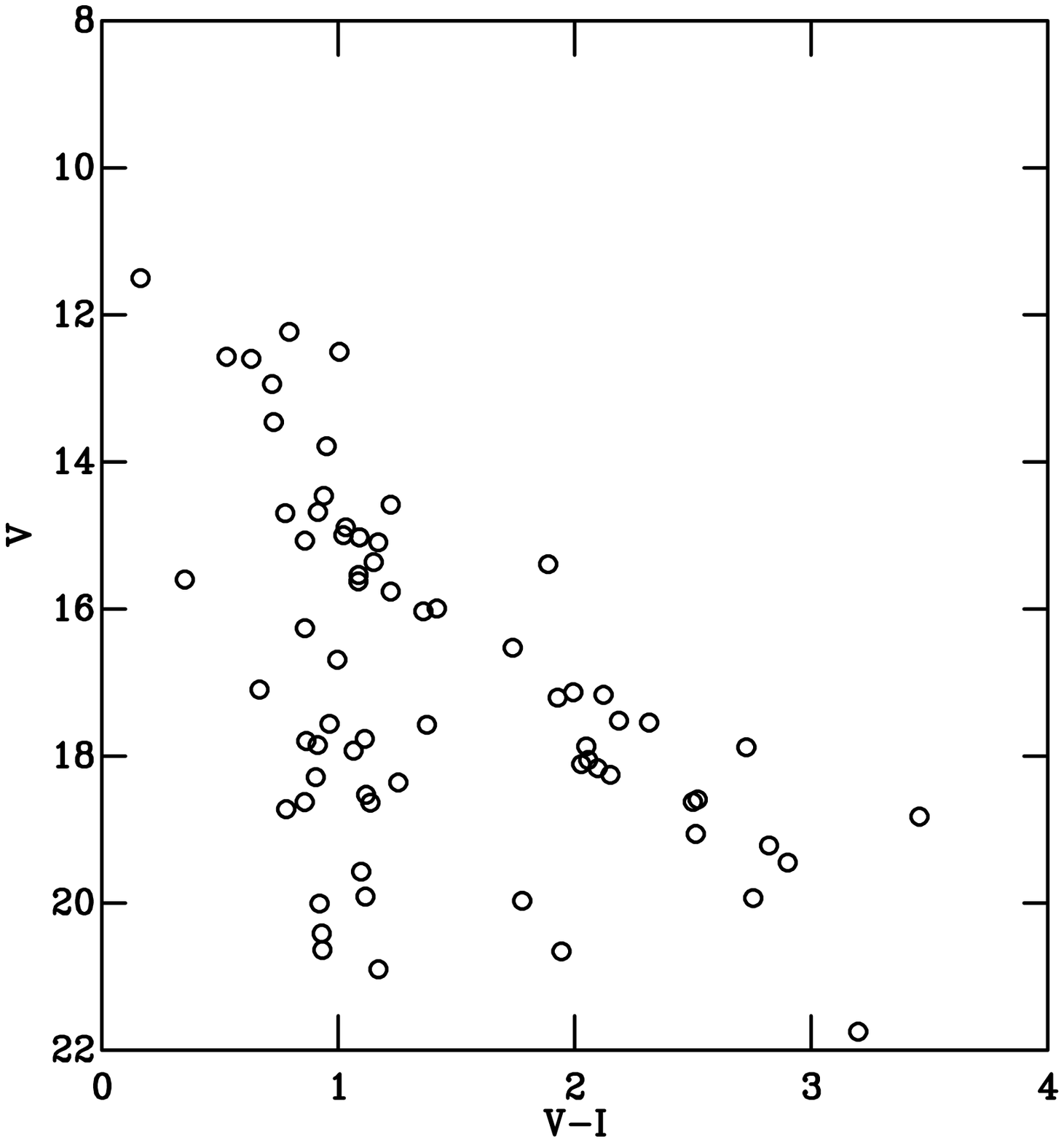}
\caption{The CMD for X-ray correlated stars.  Of the 103 stars
  correlated, 35 have been excluded from this plot due to low
  signal-to-noise (13) or for poor quality photometry (22).}
\label{fig:xrcmd}
\end{figure}

XMM-Newton source list and the reduced ROSAT PSPC source list were
cross-correlated against the optical catalogue.  XMM-Newton sources
were matched using a search radius of 5 arcseconds, except in cases
where the error in the position given in the EPIC source list was
greater than this.  In these cases the radius of the position error
circle given in the XMM-Newton data was used.  
Since the accuracy of their positions was poorer, a
search radius of 10 arcseconds was used to match the ROSAT sources.
A boresight correction of +3 arcseconds in declination was applied to
the ROSAT co-ordinates.  This was based on the offset between the
co-ordinates of the brightest X-ray source in the original ROSAT PSPC
source list and the co-ordinates of WR 6 in the optical catalogue.
  Of the 151 X-ray sources in optical
survey region (138 XMM-Newton sources and 13 ROSAT PSPC sources) 103
correlate with stars in the optical catalogue (91 XMM-Newton sources
and 12 ROSAT PSPC sources).       
The positions in colour-magnitude space of the stars in the optical
catalogue with X-ray correlations are shown in
figure~\ref{fig:xrcmd}.  We have excluded stars with data
quality flags other than ``OO'', and those with signal-to-noise less than 10.
It is clear from this plot that the majority of the correlated stars
lie the region of the diagram occupied by the PMS.  In fact, of the 68
sources plotted in figure~\ref{fig:xrcmd}, 43 lie in the PMS region
of the diagram. 
The inferred X-ray fluxes for these objects are
consistent with their status as PMS stars \citep{fcmg93}.

We estimated how many chance correlations we would expect to find in
 different regions of the CMD by performing the
 cross-correlation procedure for 8 different offsets of 30 arcseconds
 in each axis between the catalogues.    
 We find that we would expect up to 5 of the correlations in
the PMS region of the CMD to spurious, whilst essentially all of the
correlations in the background region would be spurious.

\subsection{Investigation of the proper motions of the X-ray selected
  PMS candidates}

\begin{figure}
\includegraphics[height=375pt,width=275pt, angle=90]{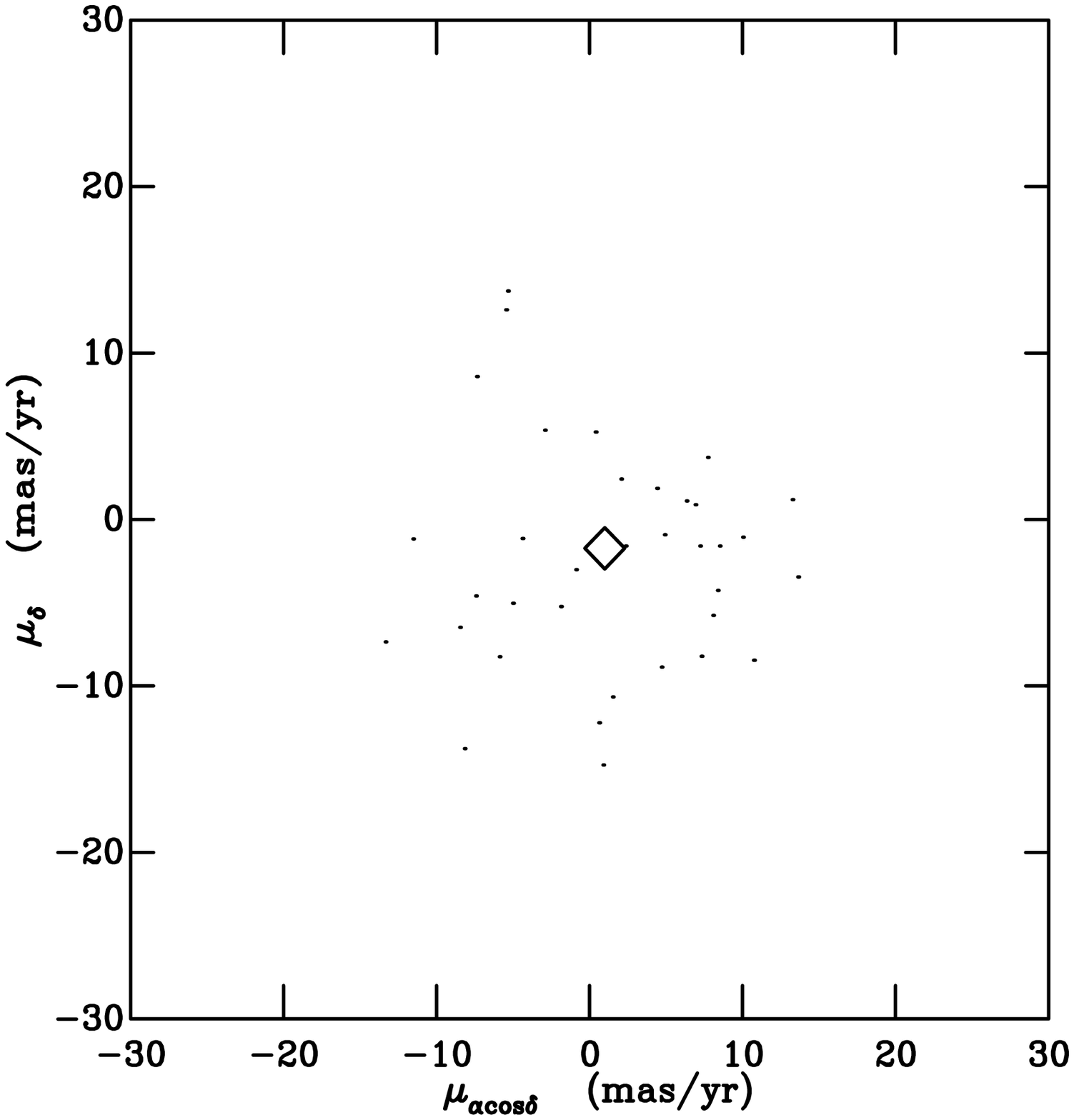}
\caption{The proper motions for our X-ray selected PMS. 
Black dots show SuperCosmos proper motions for individual stars 
\citep{supercos1}, whilst the black error diamond indicates their
weighted mean.
The mean error in the SuperCosmos proper motions is 9.7 mas/yr
in $\mu_{\alpha cos \delta}$ and 8.3 mas/yr in $\mu_\delta$.} 
\label{fig:pmfig}
\end{figure}

To remove objects that were not members of the group of interest, we
refined our selection to remove objects that do not share common motion.
  We obtained SuperCosmos proper motions for 52
stars in our X-ray selected PMS (we did not exclude stars with poor
photometry flags or low signal-to-noise for this experiment). 
A mean proper motion was calculated from the data, with each
star being given a weight inversely proportional to the square of its
error.  Stars which contributed a $\chi^2 > 4$ were removed from the
selection and the weighted mean recalculated.  This was repeated until
no stars in the sample contributed a $\chi^2$ above this value.  
In total 17 stars were removed from the sample, and these are marked
on figures~\ref{fig:isocmd1} and~\ref{fig:isocmd2} with faint circles. 
A value of $\chi^2 > 4$ was chosen as the de-selection criterion since, for
a sample of 52 stars, a 2$\sigma$ clip would be expected to remove
 2 or 3 bonafide members of the group.  
It is thought that all of the stars
removed from the sample were non-members.  The distribution of the
proper motions for the final selection are shown in
figure~\ref{fig:pmfig}. The width of the distribution is consistent
with the mean errors and an internal
velocity dispersion which is close to zero.  Those objects that were
classified as PMS members of an association along this sightline by
both photometric and proper motion criteria are listed in
table~\ref{table:xraypms_pm}.  Those that were deselected based on their
proper motions are listed in table~\ref{table:xraypms_phot}.

An interesting experiment would be to compare the proper motions of
the confirmed members of this group of PMS stars with those measured
by other surveys for stars along this sightline.  Unfortunately
systematic error in the SuperCosmos proper motions is large at galactic
latitudes $|b|$ $\leq$30$^{\circ}$ \citep{supercos3}.  
As such, SuperCosmos proper motions for stars within our survey region
with $b\approx-10^{\circ}$ will have a large systematic error. 
This is both magnitude and survey plate dependant.  

We have assessed this issue by calculating the
weighted mean of the SuperCosmos proper motions for a broad selection
of background field stars in our survey.  The proper motion we find is
($\mu_{\alpha cos \delta}$, $\mu_{\delta})$ = (2.43 $\pm$ 0.26, -2.07
$\pm$ 0.27) mas/yr.  
We calculate that the proper motion that would be expected due
to galactic rotation and solar reflex for stars at 1 kpc distance in
the direction of our survey should be approximately ($\mu_{\alpha cos
  \delta}$, $\mu_{\delta}$) = (-2, 1) mas/yr.  Since, for a random
selection of field stars, one would expect the mean proper motion to
coincide with the proper motion of the LSR and solar reflex, it is
clear that whilst the SuperCosmos proper motions are sufficiently
internally consistent for crude membership selection, they are not
suitable for drawing comparisons with proper motions obtained from
other surveys.

\addtocounter{table}{3}
\begin{table*}
\caption{The Catalogue of X-ray selected PMS stars which were also
  proper motion members. Quality flags:
  1$^{st}$ character is the quality flag for the star in the V band, the
  2$^{nd}$ is for the I band.  The meanings of the flags are: (O) O.K, (N)
  Non-stellar, (E) star too close to CCD Edge, (B) Background
  fit failed, (S) Saturated, (I) Ill determined sky, (V) Variable, (F)
  bad (Flagged) pixel, (M) negative (Minus) counts.}
\label{table:xraypms_pm}
\begin{tabular}{c c c c c c c c c c c c} 
\hline
Field No. & Index No. & $\alpha$(J2000) & $\delta$(J2000) & X & Y & V & $\sigma_V$ & Quality & V-I & $\sigma_{V-I}$ & Quality \\
\hline \hline
   1  &   283 &  06 54 28.382 & -24 00 28.33 &   523.465 &  1702.193   &  17.519  &    0.040 &  OO &   2.187  &    0.040 &  OO \\
   8  &    20 &  06 54 42.700 & -24 07 14.57 &  1690.619 &  1010.048   &  13.786  &    0.006 &  OO &   0.951  &    0.008 &  OO \\
   3  &    62 &  06 54 55.245 & -23 53 38.92 &  1278.492 &   679.337   &  14.680  &    0.006 &  OO &   0.914  &    0.008 &  OO \\
   8  &   114 &  06 54 49.326 & -24 01 45.58 &  1465.167 &   190.009   &  17.544  &    0.010 &  OO &   2.315  &    0.012 &  OO \\
   1  &   241 &  06 54 23.176 & -23 58 27.34 &   701.341 &  1400.640   &  15.763  &    0.018 &  OO &   1.222  &    0.019 &  OO \\
   1  &   217 &  06 53 44.112 & -23 56 33.98 &  2036.446 &  1119.018   &  17.868  &    0.038 &  OO &   2.049  &    0.038 &  OO \\
   1  &   103 &  06 53 53.358 & -24 02  6.55 &  1719.630 &  1947.546   &  14.996  &    0.003 &  OO &   1.022  &    0.005 &  OO \\
   1  &   142 &  06 54 17.023 & -23 51 19.34 &   911.908 &   333.942   &  17.132  &    0.009 &  OO &   1.994  &    0.012 &  OO \\
   5  &   146 &  06 54 30.847 & -24 03 21.67 &   416.553 &   414.314   &  15.993  &    0.006 &  OO &   1.417  &    0.009 &  OO \\
   1  &   270 &  06 54  4.895 & -24 00  1.90 &  1325.816 &  1636.559   &  16.527  &    0.024 &  OO &   1.738  &    0.025 &  OO \\
   5  &   155 &  06 54 24.476 & -24 03 46.10 &   634.082 &   475.192   &  17.206  &    0.008 &  OO &   1.928  &    0.011 &  OO \\
   5  &   175 &  06 54 28.227 & -24 05 50.53 &   506.052 &   785.303   &  15.538  &    0.006 &  OO &   1.086  &    0.009 &  OO \\
   5  &   377 &  06 54 11.396 & -24 04 35.61 &  1080.652 &   598.685   &  18.591  &    0.013 &  OO &   2.520  &    0.015 &  OO \\
   9  &    65 &  06 53 49.227 & -24 05 44.12 &   243.212 &   786.842   &  15.093  &    0.004 &  OO &   1.169  &    0.006 &  OO \\
   1  &   669 &  06 54 23.934 & -24 00 56.84 &   675.375 &  1773.234   &  18.256  &    0.008 &  OO &   2.151  &    0.010 &  OO \\
   2  &   610 &  06 53 23.910 & -23 58 34.80 &  1051.277 &  1415.663   &  18.623  &    0.080 &  OO &   2.500  &    0.090 &  OO \\
   2  &   128 &  06 53 26.487 & -23 51 38.56 &   963.496 &   378.178   &  17.167  &    0.033 &  OO &   2.122  &    0.033 &  OO \\
   8  &    59 &  06 54 53.249 & -24 06 16.99 &  1330.721 &   866.268   &  15.025  &    0.006 &  OO &   1.090  &    0.008 &  OO \\
   1  &   711 &  06 54 30.531 & -24 02 11.12 &   450.042 &  1958.374   &  19.059  &    0.012 &  OO &   2.512  &    0.014 &  OO \\
   1  &   421 &  06 54 20.499 & -23 52 31.25 &   792.979 &   513.154   &  19.931  &    0.031 &  OO &   2.755  &    0.033 &  OO \\
   2  &    28 &  06 53 45.803 & -24 00  4.24 &   303.310 &  1638.443   &  13.306  &    0.012 &  FO &   0.711  &    0.016 &  FO \\
   2  &   310 &  06 53 39.902 & -23 49 58.21 &   504.650 &   127.940   &  19.901  &    0.249 &  IO &   2.729  &    0.264 &  IO \\
   6  &    26 &  06 55  5.986 & -23 48  6.44 &   886.951 &  1570.648   &  12.598  &    0.008 &  OO &   0.632  &    0.009 &  OO \\
   1  &   243 &  06 54 32.055 & -23 58 39.95 &   397.960 &  1432.076   &  15.624  &    0.017 &  OO &   1.085  &    0.018 &  OO \\
   2  &   123 &  06 53 14.140 & -23 51 12.99 &  1385.775 &   314.702   &  18.053  &    0.054 &  OO &   2.057  &    0.055 &  OO \\
   1  &    50 &  06 54  4.958 & -23 51 58.68 &  1324.456 &   432.186   &  14.461  &    0.013 &  OO &   0.939  &    0.014 &  OO \\
   2  &   126 &  06 52 59.607 & -23 51 31.79 &  1882.788 &   362.056   &  17.881  &    0.048 &  OO &   2.726  &    0.048 &  OO \\
   1  &    62 &  06 53 58.401 & -23 53 50.12 &  1548.458 &   710.112   &  14.581  &    0.013 &  OO &   1.222  &    0.015 &  OO \\
   4  &   367 &  06 54  1.454 & -23 38 28.39 &  1416.782 &   120.055   &  18.109  &    0.018 &  OO &   2.028  &    0.020 &  OO \\
   4  &    18 &  06 53 59.247 & -23 42 24.56 &  1491.916 &   708.767   &  10.623  &    0.011 &  IN &   0.015  &    0.015 &  IN \\
   4  &   343 &  06 54 26.122 & -23 37 49.12 &   571.728 &    21.888   &  19.264  &    0.142 &  OO &   2.538  &    0.142 &  OO \\
   4  &   380 &  06 54 14.114 & -23 38 48.92 &   983.074 &   171.002   &  19.447  &    0.023 &  OO &   2.901  &    0.025 &  OO \\
   9  &    84 &  06 53 28.076 & -24 09 18.79 &   965.046 &  1321.824   &  14.888  &    0.006 &  OO &   1.032  &    0.008 &  OO \\
   8  &    26 &  06 54 43.492 & -24 11  0.44 &  1663.026 &  1572.877   &  11.588  &    0.011 &  NN &   0.138  &    0.016 &  NN \\
   8  &   116 &  06 54 49.690 & -24 01 54.07 &  1452.693 &   211.155   &  16.031  &    0.007 &  OO &   1.360  &    0.008 &  OO \\
\hline
\end{tabular}
\end{table*}

\begin{table*}
\caption{The Catalogue of X-ray selected PMS stars which were not
  proper motion members. Quality flags:
  1$^{st}$ character is the quality flag for the star in the V band, the
  2$^{nd}$ is for the I band.  The meanings of the flags are: (O) O.K, (N)
  Non-stellar, (E) star too close to CCD Edge, (B) Background
  fit failed, (S) Saturated, (I) Ill determined sky, (V) Variable, (F)
  bad (Flagged) pixel, (M) negative (Minus) counts.}
\label{table:xraypms_phot}
\begin{tabular}{c c c c c c c c c c c c} 
\hline
Field No. & Index No. & $\alpha$(J2000) & $\delta$(J2000) & X & Y & V & $\sigma_V$ & Quality & V-I & $\sigma_{V-I}$ & Quality \\
\hline \hline
   1  &    10 &  06 54 24.926 & -23 50 32.22 &   641.610 &   216.465   &  12.291  &    0.011 &  NN &   0.974  &    0.016 &  NN \\
   5  &    75 &  06 54 17.359 & -24 09  2.91 &   876.917 &  1264.800   &  13.457  &    0.006 &  OO &   0.727  &    0.008 &  OO \\
   1  &    73 &  06 54 11.668 & -23 56 39.46 &  1094.679 &  1131.870   &  15.363  &    0.006 &  OO &   1.150  &    0.009 &  OO \\
   1  &    54 &  06 54 38.612 & -23 52 46.11 &   173.600 &   550.238   &  14.697  &    0.005 &  OO &   0.776  &    0.006 &  OO \\
   4  &    32 &  06 54 25.237 & -23 48 13.95 &   602.029 &  1579.247   &  10.399  &    0.011 &  II &   0.530  &    0.015 &  II \\
   9  &    10 &  06 53 47.900 & -24 04 14.41 &   288.466 &   563.284   &  12.941  &    0.008 &  OO &   0.720  &    0.011 &  OO \\
   1  &   527 &  06 54 34.361 & -23 56 31.49 &   319.100 &  1111.931   &  19.216  &    0.019 &  OO &   2.822  &    0.021 &  OO \\
   5  &   474 &  06 54 15.829 & -24 07 41.69 &   929.199 &  1062.385   &  19.367  &    0.145 &  OO &   2.456  &    0.158 &  OO \\
   1  &   198 &  06 54  2.020 & -23 55 24.62 &  1424.525 &   945.530   &  18.823  &    0.014 &  OO &   3.458  &    0.016 &  OO \\
   8  &   825 &  06 54 35.539 & -24 01 26.02 &  1936.076 &   141.717   &  21.748  &    0.087 &  OO &   3.199  &    0.089 &  OO \\
   8  &    45 &  06 55  1.219 & -24 03 24.50 &  1058.879 &   436.294   &  15.070  &    0.006 &  OO &   0.859  &    0.009 &  OO \\
   4  &    25 &  06 53 47.103 & -23 44 23.71 &  1907.386 &  1006.131   &  12.502  &    0.011 &  OO &   1.005  &    0.016 &  OO \\
   2  &    51 &  06 53  1.832 & -23 51 36.62 &  1806.706 &   373.990   &  15.391  &    0.016 &  OO &   1.888  &    0.017 &  OO \\
   4  &    28 &  06 54  1.644 & -23 46 13.73 &  1409.453 &  1279.888   &  11.039  &    0.011 &  II &   1.214  &    0.015 &  II \\
   4  &    29 &  06 53 48.959 & -23 46 15.28 &  1843.567 &  1284.157   &  12.570  &    0.011 &  OO &   0.528  &    0.013 &  OO \\
   9  &    23 &  06 53 46.731 & -24 07 19.32 &   328.471 &  1024.081   &  12.231  &    0.005 &  OO &   0.793  &    0.007 &  OO \\
   5  &  3692 &  06 54  2.434 & -24 11 47.66 &  1385.935 &  1675.635   &  20.010  &    0.275 &  NI &   4.027  &    0.278 &  NI \\
   4  &   216 &  06 54 26.372 & -23 43 44.89 &   563.172 &   908.619   &  18.164  &    0.011 &  OO &   2.097  &    0.014 &  OO \\
\hline
\end{tabular}
\end{table*}

\section{Discussion: How far away and how old are the PMS stars?}

The fitting of theoretical isochrones to the observed PMS can yield
distances and ages for stars found there.  However there is a
degeneracy between distance and age for low-mass stars.  
As such any estimates of distance and
age must be made within the context of the estimates of the
distances to the higher mass stars, as well as the constraints that
higher mass stars place on the age of any associations that may be
present.

Isochrones in $V/V-I$ were generated from the solar metallicity models
of \citet{dam97} in the same manner as described by \citet{jth2001}. 
Empirical colour-$T_{eff}$ conversions were obtained by fitting a 120 Myr
isochrone to the Pleiades at a distance modulus of 5.6, $E(B-V)$=0.04, $E(V-I_{c})$=0.05. 

\begin{figure}
\includegraphics[height=375pt,width=275pt, angle=90]{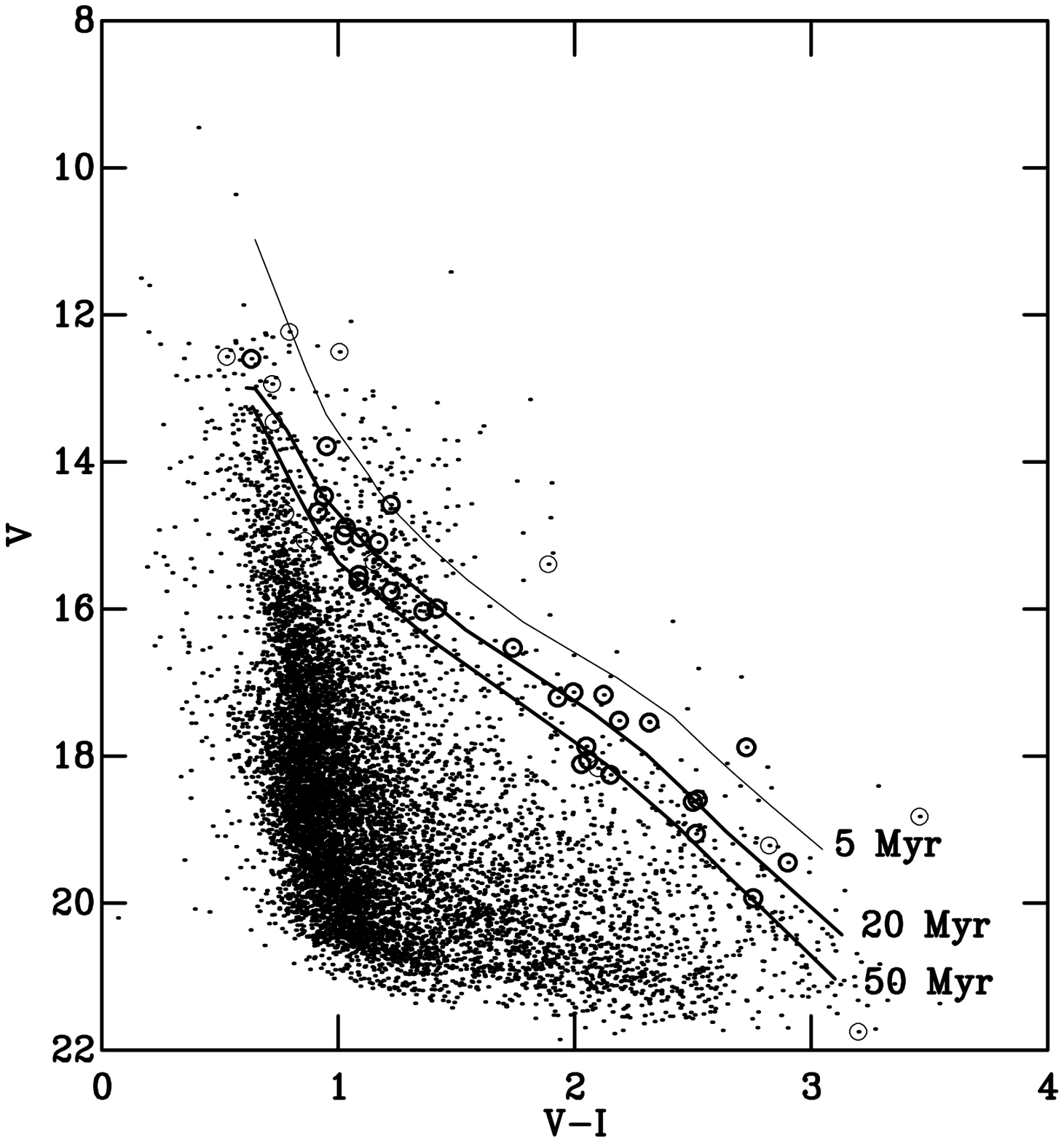}
\caption{The CMD for the fields centred on WR 6 with 5, 20, and 50
  Myr isochrones overlaid for (intrinsic) dM=8.9 and $A_V$=0.17. Circled points
  indicate X-ray correlated stars.  Points marked with faint circles are
  those cut from the sample in section 4.2.}
\label{fig:isocmd1}
\end{figure}

We scaled the resultant isochrones for a distance of 600 pc, in
keeping with the distance derived for Cr 121 by \citet{zhbbb99} of 592
$\pm$ 28 pc.  In addition to shifting the isochrones for distance,
they were also shifted for an extinction of $A_V = 0.17$
\citep{k2000}.  This corresponds to a reddening of  $E(V-I) = 0.07$
(if we assume a galactic reddening law),
though this has little effect on the results as the reddening vector
lies along the isochrones. 
  If the low-mass PMS stars are located within the
association at the distance found from the Hipparcos data, the
best fitting isochrones are those corresponding to ages between 20 and
50 Myrs (see figure~\ref{fig:isocmd1}).  These imply a median age of 30
Myrs, with an age spread of about 30 Myrs.  Such an age spread is not
considered likely for an association of this age \citep[][and
references therein]{jt98}.
  Additionally, as was noted by
\citet{zhbbb99}, the presence of an O-star and  early B-stars
indicates that this moving group is young, about 5 Myrs old.    
For such a large number of PMS stars to be found within such a small
survey region would also be surprising if they were associated
with the group described by \citet{zhbbb99}.  
Assuming a Salpeter initial mass function (IMF), we calculate that we
would expect to find 3900 0.20-0.73 M$_{\sun}$ ($14\leq V \leq 20$ at
592 pc)
stars for the 85 B stars selected as members by \citet{zhbbb99}.
If we assume that these would be distributed evenly over a region at
least as large as the 13$^{\circ}$x16$^{\circ}$ region occupied by
the B stars, then we would expect to find fewer than five 0.20-0.73 M$_{\sun}$ stars within
the region observed here.  As such, the PMS
found here would represent quite a concentration of the low-mass stars
of the association.  It should be noted
that a more realistic IMF would predict even fewer stars in this mass
range, since several observational studies have found that the IMF in
young clusters flattens out at masses below $\sim$0.8 M$_{\sun}$
\citep[e.g.][and references therein]{pbbgz2002,blhsk2002}.

\begin{figure}
\includegraphics[height=375pt,width=275pt, angle=90]{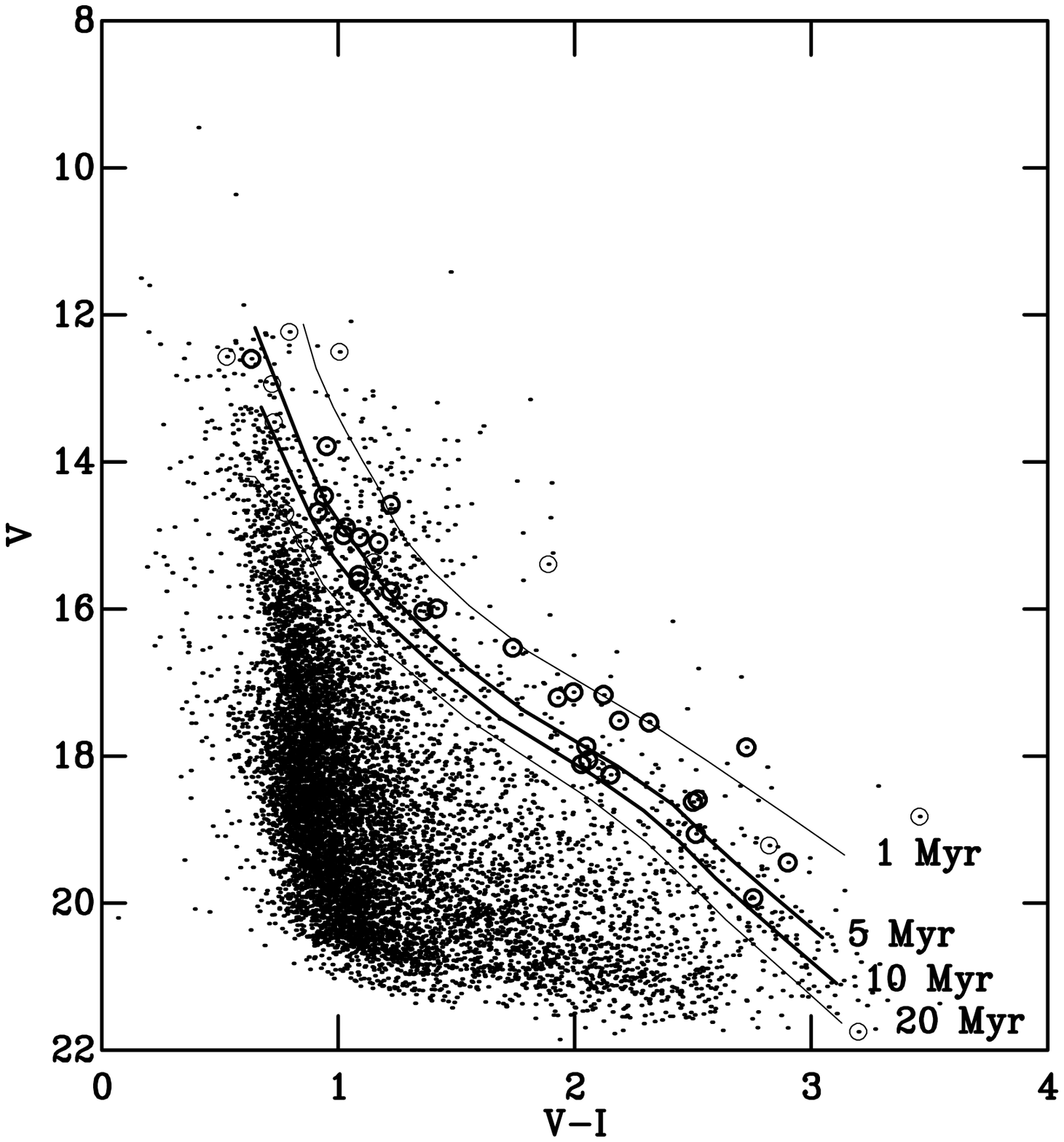}
\caption{The CMD for the fields centred on WR 6 with 1, 5, 10 and 20
  Myr isochrones fitted for (intrinsic) dM=10.1 and $A_V$=0.17. Circled points
  indicate X-ray correlated stars.  Points marked with faint circles are
  those cut from the sample in section 4.2.}
\label{fig:isocmd2}
\end{figure}

The three lines of argument described above lead us to reject the
scenario in which the low-mass PMS stars are associated with the
moving group detected using the Hipparcos data.  Instead, we use the
presence of early-type main sequence B-stars in the membership lists
for both interpretations of Cr 121 to constrain the age to less than
15 Myrs.  Such a young age is also more in keeping the scatter of
stars in the PMS region of the diagram.
Scaling sets of isochrones for ages of less than 15 Myrs for distance
until they fitted as much of the data as possible gave a distance of
1050 pc, and a modal age of between 1 and 10 Myrs, applying the same
extinction as before.    
This is consistent with the distance given by \citet{k2000} for a compact
group of stars at 1085 $\pm$41 pc.  Figure~\ref{fig:isocmd2} shows
that the age spread implied by fitting isochrones for this distance is
less than 10 Myrs, much more in keeping with age spreads observed
in other young clusters and associations \citep[e.g.][]{pnjd2003}, than
  the 30 Myr age spread suggested by the fit for 600 pc.
Isochrones for 1 and 20 Myr, scaled for the same distance as the 5 and
10 Myr ones, are included in figure~\ref{fig:isocmd2} for comparison.
  
If we carry out a similar calculation to the one above,
using a Salpeter IMF, we find that we would expect there to be around 540
stars in the 0.29-0.98 M$_{\sun}$ mass range ($14\leq V \leq 20$ at
1050 pc) for the 11 B-type members
of the more distant cluster suggested by \citet{k2000}. If we assume an even
distribution of stars over the $\approx$1$^{\circ}$ radius surveyed
there, then we would expect to find, at most, 130 stars in this mass range
in the region covered by our optical survey.  This is consistent with
the population of the PMS observed here.
This would indicate that the
PMS we have identified consists of stars in the $\approx$ 1 kpc
distant cluster identified by \citet{c31,e81} and \citet{k2000}. To
avoid confusion we shall refer to this
distant cluster as Cr 121, and we shall refer to the association at
approximately 600 pc as CMa OB2, following \citet{e81}.

\section{Conclusions}

It seems clear that the low-mass PMS stars detected here are
associated with the compact group of stars found by \citet{k2000,e81}
and \citet{c31} at a distance of over 1 kpc, originally designated Cr
121.  If this cluster were of the extent and at the distance found by
\citet{zhbbb99}, the discovery of such a concentration of low-mass PMS
stars which are so much older than the rest of the association, and
with an age spread of 30 Myrs, seems incredible.  It is also
clear however, that there is a young moving group that was detected using the
Hipparcos data by \citet{zhbbb99} at a distance of 592 $\pm$28 pc,
of which WR 6 is likely a member, in the same direction.  The
characteristics of the group described there, however, seem more in keeping
with an OB association, than with the compact open cluster originally
described as Cr 121.  We argue that the distant open cluster
should retain its original designation as Collinder 121, whilst the
nearer OB association described by \citet{f67,zhbbb99,daml2002} should
be re-designated as CMa OB2, following \citet{e81}.  

Whilst the Hipparcos census of OB associations within 1 kpc is an
invaluable resource for studying recent local star formation, this work
demonstrates the importance of interpreting proper-motion and parallax
data within the context of age and distance constraints imposed by
main-sequence high mass and low-mass PMS stars.  This is
particularly true when investigating structures that extend beyond
the range of the Hipparcos data.

\section*{Acknowledgements}

We thank Nigel Hambly for useful comments regarding the use of proper
motions from the SuperCosmos sky surveys, and Mike Watson for alerting
us to the existence of the XMM-Newton data for WR 6 and S308.
  We also thank our referee
Ronnie Hoogerwerf for many useful comments which improved this paper, and
for alerting us to problems in our use of the SuperCosmos
proper motions.  
The Cerro Tololo Interamerican Observatory
is operated by the Association of Universities for Research in
Astronomy, Inc., under contract to the US National Science Foundation.
This work made use of ROSAT data obtained from the Leicester Database
Archive Service at the Department of Physics and Astronomy, Leicester
University.

\bibliography{cr121refs}
\bibliographystyle{mn2e}

\end{document}